\documentclass{iaus}
\usepackage{graphicx}

\usepackage{epsf}
\usepackage{amsmath}
\allowdisplaybreaks

\newcommand{\Teff}{\mbox{$T_\mathrm{eff}$}}
\newcommand{\K}{\,{\mathrm K}}
\newcommand{\der}{\mathrm d}
\newcommand{\Lamastnu}{\mbox{$\Lambda^\ast_\nu$}}

\newcommand{\tentosbornik}{these proceedings}

\DeclareMathAlphabet{\mathsc}{OT1}{cmr}{m}{sc}
\def\testbx{bx}%
\DeclareRobustCommand{\ion}[2]{%
\relax\ifmmode
\ifx\testbx\f@series
{\mathbf{#1\,\mathsc{#2}}}\else
{\mathrm{#1\,\mathsc{#2}}}\fi
\else\textup{#1\,{\mdseries\textsc{#2}}}%
\fi}

\title{Standard model atmospheres for A-type stars and non-LTE effects}

\author[J.~Kub\'at, D.~Kor\v{c}\'akov\'a]
{Ji\v{r}\'{\i} Kub\'at$^1$ \and Daniela Kor\v{c}\'akov\'a$^2$}

\affiliation{Astronomick\'y \'ustav,
Akademie v\v{e}d \v{C}esk\'e republiky,
CZ-251 65 Ond\v{r}ejov,
Czech Republic
\break
email: $^1$kubat@sunstel.asu.cas.cz,
       $^2$kor@sunstel.asu.cas.cz
}

\pubyear{2004}
\volume{224}  
\pagerange{1--10}
\date{?? and in revised form ??}
\jname{The A-Star Puzzle}
\editors{J.\,Zverko, W.W.\,Weiss, J.\,\v{Z}i\v{z}\v{n}ovsk\'{y}, \& S.J.\,Adelman, eds.}
\begin{document}

\maketitle

\begin{abstract}
The current status of NLTE model atmosphere calculations of A type stars is
reviewed.
During the last decade the research has concentrated on solving the 
restricted NLTE line formation problem for trace elements assuming LTE
model atmospheres.
There is a general lack of calculated NLTE line blanketed model
atmospheres for A type stars, despite the availability of 
powerful methods and
computer codes that are able to solve this task.
Some directions for future model atmosphere research are suggested.
\end{abstract}

\firstsection 
\section{Introduction}

Compared to the atmospheres of hotter O and B stars that have stellar winds
and to atmospheres of cooler F and G stars that have convective
atmospheres, the atmospheres of A-type stars are relatively quiet,
which enables the presence of various interesting phenomena.
This calmness of atmospheres leads to the development of a number of
different chemical peculiarities and magnetic field structures on long
time scales.

There exists a widely used grid of line blanketed LTE model
atmospheres (Kurucz \cite{kurmod}, Castelli \& Kurucz
\cite{novyatlas9}), which is also extensively used for A stars.
Besides this grid, several attempts to remove the inconsistent
assumption of LTE have been performed.
The very first
NLTE calculation of a cool B star, which can be somehow
understood as a first step toward A star modelling, was performed by
Auer \& Mihalas (\cite{AM70}).
The domain of A star temperatures was first reached by Kudritzki
(\cite{A0Ku}) and Frandsen (\cite{A0Fr}).
A grid of NLTE model atmospheres of stars with $10000\K \le \Teff \le
15000\K$ (thus also including a part of the A star domain) was
calculated by Borsenberger \& Gros (\cite{ABoGr}).
A great stride was made by Huben\'y (\cite{AHu80}, \cite{AHu81}), who
studied in detail the effect of the L$\alpha$ line on the NLTE model
atmospheres of A stars and applied his code to star Vega.
A detailed analysis of the physics which is contained in releasing the
assumption of LTE was presented by Huben\'y (\cite{CPrev}).
First NLTE line blanketed model atmospheres of A stars were calculated
using a method of superlevels and superlines by Hubeny \& Lanz
(\cite{revsup}).
Unfortunately, although there was great progress in calculating 
NLTE model atmospheres for hot stars and for white dwarfs,
calculations of NLTE model atmospheres
of A stars
are very rare.
The only attempt of a NLTE model atmosphere of an A-type star was the
model calculated by Hauschildt et al. (\cite{phoenix}) for parameters
corresponding to Vega.
The authors announced a grid of NLTE models to be published, but till
now nothing appeared.

The inherent reason which stands tacitly behind the lack of calculated
NLTE A star model atmospheres is in the different scaling properties of the
atmospheres of this stellar type.
The atmosphere of A stars have very extended line formation regions
meaning that different lines form at very different depths (cf.
Fig.\,1 in Lanz \& Hubeny, \cite{diag93}). Therefore they are very difficult to
describe by a simple standard depth discretization as compared to O
stars and white dwarfs, and hence a much more elaborate depth scale has to be
used.
Our own experience shows that there are severe convergence problems
caused by improper depth scaling and only a tremendous number of depth
points helped.

\section{Basic equations for NLTE model atmospheres}

A solution of a model atmosphere problem means taking the basic input values
of stellar luminosity $L$, stellar radius $R$, and stellar mass $M$ (or,
equivalently for the usual plane-parallel case effective temperature
{\Teff} and gravitational acceleration at the stellar surface $g$),
adding physical laws that are important in the stellar atmosphere (a principal
ingredient),
and calculating the space distribution of
various physical quantities (temperature $T(\vec{r})$, electron density
$n_e(\vec{r})$, population numbers $n_i(\vec{r})$, density
$\rho(\vec{r})$, velocity field $\vec{v}(\vec{r})$, etc.).
This procedure is often being performed in substeps.
The first step is the determination of $T$, $n_e$, $\rho$, and $n_i$ for
the most important opacity sources.
The second (optional) step is the
determination of $n_i$ for the minor
abundant species, often referred to as trace elements.
The final step in this process is the calculation of the theoretical
emergent radiation which may be then compared to the observed one.

A standard model atmosphere here is a static one-dimensional
(plane-parallel or spherically symmetric) atmosphere in hydrostatic,
radiative, and statistical equilibrium.
To calculate such a model atmosphere we need to solve the radiative
transfer equation (to determine the radiation field $I$),
the hydrostatic equilibrium equation (to determine $\rho$), the energy
equilibrium equation (to determine $T$), and the statistical
equilibrium equations (to determine $n_i$).

\subsection{Radiative transfer equation}

The radiative transfer equation is the principal equation for model
atmosphere calculations.
Regardless of the geometrical approximation which is adopted, the basic
solution of the radiative transfer equation is performed along a ray,
which reads
\begin{equation}\label{rteray}
{\der I_{\nu} (r) \over \der s} =
- \chi_\nu (r) I_{\nu} (r)
+ \eta_{\nu} (r). 
\end{equation}
Usually the simplest possibility, i.e. the plane-parallel atmosphere, is
being used for the modelling of stellar atmospheres, basically due to its
relative simplicity.

\subsubsection{Formal solution}

By formal solution of Equation (\ref{rteray}) we mean the solution
for a {\em given} opacity and emissivity.
Such a solution may be performed using either differential
or integral methods.
The latter are especially useful for the multidimensional case.
The inconsistency of the formal solution is that opacity
$\chi_\nu$
and emissivity
$\eta_{\nu}$
are not given,
but in the process of solving the model atmosphere
problem they depend on temperature, density, and radiation field, which
are to be determined when solving the radiative transfer equation.
However, the formal solution remains at the heart of each model atmosphere
code (see Auer \cite{formau}).

\subsubsection{Approximate solution using ALO}

The process of the formal solution may be expressed using the
$\Lambda$-operator
\begin{equation}
J_\nu = \Lambda_\nu S_\nu.
\end{equation}
This expression may be used in an iterative process, where the source
function $S_\nu$ is determined by solving the other constraint
equations of hydrostatic, energy, and statistical equilibria.
As has been shown, such a process, however effective for the optically
thin cases of molecular clouds (cf. Dickel \& Auer \cite{diauer}), fails
to produce a convergent solution for the optically thick
cases of stellar atmospheres (see Auer \cite{A84}).
This problem has been overcome using the Newton Raphson method for
model stellar atmospheres by Auer \& Mihalas (\cite{lin}).
As an efficient alternative, an approximate $\Lambda$-operator
{\Lamastnu} may be used,
\begin{equation}
J_\nu = \Lamastnu S_\nu + \left( \Lambda_\nu - \Lamastnu \right) S_\nu
\end{equation}
The operator {\Lamastnu} is constructed in such a way that it contains the
basic physics of the problem and is simply calculable.
The most efficient method is to calculate this operator consistently
with the numerical method used for the formal solution, like it has been
described by Rybicki \& Hummer (\cite{RH91}) or Puls (\cite{P91}).
Then the radiation field is determined by an iterative process,
\begin{equation}\label{aliiter}
J_\nu^{(n)} = \Lamastnu \left[ S_\nu \left( J_\nu^{(n)} \right)
\right]
+ \left( \Lambda_\nu - \Lamastnu \right)
\left[ S_\nu
\left( J_\nu^{(n-1)} \right) \right]
\end{equation}
where $(n)$ means the current iteration step and $(n-1)$ the previous one.

\subsection{Energy equilibrium}

The energy equilibrium equations determine the temperature
structure.
They describe the basic energy balance in the atmosphere.
For the case of static atmospheres, the equation of radiative
equilibrium,
\begin{equation}
\int_0^\infty \left( \kappa_\nu J_\nu - \eta_\nu \right) d \nu =
\int_0^\infty \kappa_\nu \left( J_\nu - S_\nu \right) d \nu  = 0,
\end{equation}
which simply describes the radiative energy balance, is
usually being used.
However, this equation does not guarantee the radiative flux
conservation, which results in an incorrect temperature structure at large
optical depths.
Therefore, another form, which comes from
$\nabla \cdot \vec{\cal{F}} =0$, is to be used at large optical depths.
Better numerical properties of the scheme may be achieved if a linear
combination of these equations is considered, as was first done by
Hubeny \& Lanz (\cite{nltesuper}).
Another improvement of convergence properties, especially useful at low
continuum optical depths where strong lines are still present, was
introduced by Kub\'at et al. (\cite{kpptt}) where
a thermal balance of electrons instead of the radiative equilibrium
equation is used.

Then three different equations may be used in the stellar atmosphere to
take into account energy equilibrium.
In the deepest layers, the differential form of radiative equilibrium
is used.
In the middle parts, where continuum radiation is formed, the integral
form of radiative equilibrium works well.
In the outer parts where optically thick lines coexist with an optically
thin continuum (which is the case of A star atmospheres), the thermal
balance of electrons works best.
A more detailed discussion with the numerical properties of these methods
may be found in Kub\'at (\cite{tuetb,ATAmod}).
 
\subsection{Equations of statistical equilibrium}

The equations of statistical equilibrium are the key equations both for
the NLTE model atmosphere and line formation problems.
The set of equations for statistical equilibrium for the static case
may be written as
\begin{equation}\label{ese}
n_i \sum_l (R_{il} + C_{il}) - \sum_l n_l (R_{li} + C_{li}) = 0
\end{equation}
Radiative rates $R_{il}$ are those responsible for the NLTE effects,
since they introduce nonlocal interaction and they alter the level
population irrespective of local equilibrium conditions.
They move the gas outside thermodynamic equilibrium.
On the other hand, for an equilibrium (Maxwellian) electron velocity
distribution (which is the common case), collisional rates reintroduce
thermodynamic equilibrium.
The balance between collisional and radiative rates determines the
applicability of the LTE approximation.
If collisions dominate, then the LTE approximation is feasible.
If radiative transitions dominate, which is the case in A star
atmospheres, then equations of statistical equilibrium need to be solved
if we want to obtain the correct population numbers.

Since the system of rate equations is linearly dependent, we have to
close it with some other condition.
For a reference (usually, but not necessarily hydrogen) atom either
particle conservation equation
$\sum_k N_k = N - n_e$
or charge conservation equation
$\sum_k \sum_j q_j N_{jk} = n_e$
are used.
For other atoms, the abundance equation
$N_k = Y_k N_r$
is usually used.
Detailed forms of the equations may be found, e.g., in Kub\'at
(\cite{ATA3}).

\subsection{Solution of the system of equations}

The whole system of equations is usually solved using a Newton-Raphson
iteration scheme, which in modeling stellar atmospheres used to be referred to
as linearization (Auer \& Mihalas \cite{lin}).
The radiation field is linearized as well, or it can be included
using approximate lambda operators (see
Eq.~\ref{aliiter}), as has been first done by Werner (\cite{aliklaus}).
Excellent reviews of ALI methods of model atmosphere solutions are provided by
Hubeny (\cite{alirev1}) and Werner {\etal} (\cite{alirev2}).
In addition, an implicit linearization of $b$-factors saves additional
computer time and memory (Anderson \cite{block}).
The whole process may be accelerated using either Kantorovich (see
Hubeny \& Lanz \cite{kantor}) or Ng acceleration (see Auer
\cite{accel}).

\subsection{Line blanketing}

Line blanketing is caused by enormous amounts of spectral lines in the UV
region ($\sim 10^7$), mostly of iron and nickel.
This enormous amount of lines causes radiation to be absorbed in the UV
and reemitted in the visible region, thus changing the emergent flux
and atmospheric structure significantly.

There are two main approaches to line blanketing under the
simplifying assumption of LTE, the ODF (Opacity Distribution
Function) and OS (Opacity Sampling). 

If we do not use the assumption of LTE, we have to solve the equations
of statistical equilibrium to obtain the correct population numbers for all
levels.
This task is relatively simple for simple atoms (like He), but becomes
difficult for atoms with a complicated level structure, like iron.
In order to cope with the complexity of these atoms, Anderson
(\cite{super}) introduced the concept of superlevels and superlines.
A superlevel is a level which is created by grouping several individual
levels.
In order to achieve a real simplification, it is useful to group the
levels in such a manner that the relative population distribution inside the
superlevel obeys the LTE distribution.
Thus, this grouping is done according to the energy, $E_i$, of
individual levels (Anderson \cite{super}, Dreizler \& Werner
\cite{wdblanket}).
A more sophisticated grouping according to $E_i$ and parity was done by
Hubeny \& Lanz (\cite{nltesuper}).
Dreizler \& Werner (\cite{wdblanket}) used opacity sampling in their
calculations, whereas Hubeny \& Lanz (\cite{nltesuper}) used the NLTE
opacity distribution function.

\subsection{LTE versus NLTE}

Since the beginning of the NLTE calculations there is a battle between
'LTE people' and 'NLTE people', which of the approaches is better.
If NLTE model atmosphere is calculated as easily as the LTE one, there
would be no discussion and everybody would use the NLTE one.
However, this is not the case and it is much more difficult to calculate a
NLTE model than the LTE model.
Two conflicting aims enter the scene.
The first one is to analyse as many stars as possible, which can be
hardly done using NLTE models.
The second one is to study stars as accurately as possible, which can be
hardly done using LTE models.

For the case of LTE it is relatively easy to handle line blanketing,
since one neglects the effect of radiation on atomic population numbers,
which saves computing time enormously.
On the other hand for a more general case of NLTE, including line
blanketing is computationally expensive, albeit nowadays it is possible
to handle NLTE calculations using sophisticated numerical methods and contemporary
computers.
However, for extended atmospheres the reliability of LTE decreases and
one is forced to use NLTE.

To avoid repeated calculations of complicated model atmospheres one may
take the advantage of precomputed grids.
However, one has to keep in mind that using such grids, which were
calculated under certain physical assumptions, always limits the results
by these assumptions.

\section{NLTE line formation calculations}

\begin{table}
\caption{List of NLTE calculations for a given LTE model atmosphere for
A stars}\label{resnlt}
\begin{tabular}{lllp{184pt}}
\hline
Ion & reference & code & comment on included levels and transitions\\
\hline
\ion{He}{i} & Takeda (\cite{hedeneb}) & 3
   & calculations for Deneb; 88 levels \\
\ion{Li}{i} & Mashonkina {\etal} (\cite{mashli}) & 4 & \\
\ion{Li}{i/ii} & Shavrina {\etal} (\cite{litatari}) &
   & 20 levels \\
\ion{C}{i} & Takeda (\cite{ytcn}) & 3
   & 129 levels, 2351 radiative transitions \\
           & Venn (\cite{V95}) & 2
   & 83 levels \\
           & Rentzsch-Holm (\cite{inga96b}) & 2
   & 83 levels, 63 transitions \\
           & Paunzen {\etal} (\cite{P99}) & 2
   & 83 levels, 63 transitions \\
\ion{C}{i/ii} & Przybilla {\etal} (\cite{PBK01}) & 1
   & \ion{C}{i} levels with $n\le 9$,
     \ion{C}{ii} levels with $n\le 4$ \\
\ion{N}{i} & Takeda (\cite{ytcn}) & 3
   & 119 levels, 2119 radiative transitions \\
           & Sadakane {\etal} (\cite{ndeneb}) & 3
   & used model atom of Takeda (\cite{ytcn}) \\
           & Takeda \& Takada-Hidai (\cite{ytn1s1}) & 3
   & used model atom of Takeda (\cite{ytcn}) \\
           & Takada-Hidai \& Takeda (\cite{ytnscp}) & 3
   & used model atom of Takeda (\cite{ytcn}) \\
           & Venn (\cite{V95}) & 2
   & 93 levels, 189 transitions \\
           & Lemke \& Venn (\cite{LV96}) & 2
   & 93 levels, 189 radiative transitions \\
           & Rentzsch-Holm (\cite{inga96a}) & 2
   & 96 levels, 82 transitions \\
\ion{N}{i/ii} & Przybilla \& Butler (\cite{PB01}) & 1
   & \ion{N}{i} levels with $n\le 7$,
     \ion{N}{ii} levels with $n\le 6$ \\
\ion{O}{i} & Takeda (\cite{yto1,yto1pop2}) & 3
   & 86 levels, 294 radiative transitions \\
           & Takeda \& Takada-Hidai (\cite{yt26}) & 3
   & 86 levels, 294 radiative transitions \\
           & Takeda et al. (\cite{ythgmn}) & 3
   &
%
     calculations from Takeda (\cite{yto1pop2}) for CP
     stars \\
           & Paunzen {\etal} (\cite{P99}) & 2
   & 15 levels \\
           & Przybilla {\etal} (\cite{Prz00}) & 1
   & all levels below excitation energy \\
\ion{Na}{i} & Takeda \& Takada-Hidai (\cite{ytna1}) & 3
   & calculations for A supergiants, 92 levels, 178 radiative
     transitions \\
            & Mashonkina {\etal} (\cite{MSS00}) & 4
   & solution for $\Teff = 4000-12\,000\K$, 21 levels \\
\ion{Mg}{i} & Shimanskaya {\etal} (\cite{SMS00}) & 4
   & solution for $\Teff = 4500-12000\K$, 50 levels \\
            & Idiart \& Th\'evenin (\cite{IT00}) & 5
   & 104 levels, 980 radiative transitions \\
\ion{Mg}{i/ii} & Przybilla {\etal} (\cite{PBBK01}) & 1
   & all levels with $n\le 10$ \\
\ion{S}{i} & Takeda \& Takada-Hidai (\cite{ytn1s1}) & 3
   & 56 levels, 173 radiative transitions \\
           & Takada-Hidai \& Takeda (\cite{ytnscp}) & 3
   & 56 levels, 173 radiative transitions \\
\ion{K}{i} & Ivanova \& Shimanskii (\cite{IS00}) & 4
   & solution for $\Teff = 4000-10\,000\K$, 36 levels \\
\ion{Ca}{i} & Idiart \& Th\'evenin (\cite{IT00}) & 5
   & 84 levels, 483 transitions\\
\ion{Ti}{ii} & Becker (\cite{sylvia}) & 1
   & complete model atom, using superlevels \\
\ion{Fe}{ii} & Becker (\cite{sylvia}) & 1
   & complete model atom, using superlevels \\
\ion{Fe}{i/ii} & Rentzsch-Holm (\cite{inga96b}) & 2
   & 79+20 levels, 52+23 lines \\
                           & Th\'evenin \& Idiart (\cite{TI99}) & 5
   & 256+190 levels, 2117+3443 lines \\
\ion{Sr}{ii} & Belyakova {\etal} (\cite{srbel}) & 4
   & solution for $\Teff = 4000-12\,000\K$, 41 levels \\
\ion{Nd}{ii/iii} & Mashonkina {\etal} (\cite{mashnd}) & 1
   & 247+69 levels \\
\hline
\end{tabular}

Computer codes used for calculations:
1 -- DETAIL (Giddings \cite{detaildis});
2 -- Kiel (Steenbock \& Holweger \cite{Kiel});
3 -- Takeda (\cite{japonec});
4 -- NONLTE3 (Sakhibullin \cite{nonlte3});
5 -- MULTI (Carlsson \cite{multi});
\end{table}

Calculations of full NLTE model atmospheres of A type stars is a
difficult task.
Due to a large span of line formation regions down to optical depths in
the continuum of about $10^{-9}$, the numerical procedure becomes extremely
unstable and calculations need special care.
That is why people started to solve easier task of NLTE line formation
for a given model atmosphere.
A necessary condition for such calculation to be reasonable, is
negligible influence of the ion on the ionization balance and,
consequently, on the global structure of the atmosphere.
Such elements are being referred to as trace elements.

Contemporary NLTE analysis of A stars are mainly concerned with solving the
statistical equilibrium equations for trace elements for a given
(mostly LTE) model atmosphere.
These kind of calculations have already been reviewed by Huben\'y
(\cite{CPrev}) and Hubeny \& Lanz (\cite{revsup}), so only new
calculations that appeared after the last review are listed in 
Table~\ref{resnlt}.
We did our best to mention all calculations.
In the case we unintentionally omitted some, we apologize for it.

There are several codes available for this purpose.
They use slightly different numerical techniques, basic differences are in
the treatment of the radiation field.
Both the accelerated lambda iteration and complete linearization methods are
used.
A comparison of results from different codes for Vega (and Sun) was
presented by Kamp et al. (\cite{ingaupp}), who found large differences
in the results different codes.
Therefore we added to Table~\ref{resnlt} an indication which code
was used.


\section{Beyond standard model atmospheres}

\subsection{Nonthermal collisional rates}

In standard NLTE calculations only the radiation field and level
populations are allowed to deviate from their equilibrium values.
The velocity distribution of individual species is still assumed to be
in equilibrium.
Collisional rates are then calculated as an average over the
Maxwellian electron velocity distribution,
\begin{equation}
C_{ij} = n_e \int_{v_0}^\infty \sigma_{ij}(v) f(v) v \der v.
\end{equation}
Thus in equilibrium there holds a relation between collisional rates up
and down (excitation/deexcitation or ionization/recombination),
\begin{equation}\label{eqcij}
n_i^\ast C_{ij} = n_j^\ast C_{ji}.
\end{equation}
If the velocity distribution is not Maxwellian (e.g. in electron beams
formed in the Sun during flares -- cf. Ka\v{s}parov\'a \& Heinzel
\cite{janca}),
then the equation (\ref{eqcij}), which expresses the equilibrium
condition, is not valid.
In such a case the collisions may cause level population
numbers to differ from the LTE values, which has an observable effect
on the line profiles.
The nonthermal collisional term may be considered as another source of
NLTE effects.
Such effects are present in the Sun, where the electron beams are formed
after magnetic reconnection.
There is also a possibility that they may be present in magnetic
A stars, where releasing magnetic energy in eruptive events like
flares may be expected as well.

\subsection{Magnetic fields}

Inclusion of magnetic fields into model atmosphere calculations have
been done only occasionally.
An important attempt in calculating a model atmosphere with a magnetic
field was done by Carpenter (\cite{carpenter}),
who found changes in the net gravity and pressure distribution
due to the magnetic field.
The most consistent model so far was recently
developed by Valyavin {\etal} (\cite{rusi}),
who assumed LTE and
included the Lorentz force into the equation of hydrostatic
equilibrium.
A historical overview of magnetic model atmosphere calculations is
also presented there.

\subsection{Diffusion}

An important aspect of A star atmospheres is the fact that atmospheres
of a good fraction of stars of this stellar spectral type are extremely
quiet.
Such quiet atmospheres without significant global motions like global
convection and stellar winds enable long time scale processes of
diffusion and gravitational settling to take place.
Diffusion occurs not only in the presence of magnetic field
(note the effect of enhancing radiative acceleration in polarized
radiation field and the effect of ambipolar diffusion, Babel \& Michaud
\cite{babel1,babel2}),
but also due
to different sensitivities of various atoms and ions to incoming
radiation.
Radiative diffusion is able to explain various chemical peculiarities in
CP stars and is also responsible for the isotopic shift effect (Aret
\& Sapar \cite{anna}).
However, consistent NLTE model atmospheres that take into account
radiative diffusion processes are still missing for A type stars.
NLTE model atmospheres with diffusion were calculated by Dreizler \&
Shuh (\cite{sona}) for white dwarfs, where diffusion
processes are also important.
The physical process of diffusion is reviewed in detail elsewhere in
these proceedings (Michaud).

\subsection{Extended atmospheres}

Plane-parallel model atmospheres are also being used for NLTE modeling
of atmospheres of A-type supergiants (e.g., Kudritzki \cite{A0Ku}).
However, they do not describe the limb darkening correctly.
A-type supergiants have also
stellar winds and show P Cyg line profiles,
therefore it is impossible to describe them using plane
parallel atmospheres.
Also, wind models of A stars are not calculated too often.
Multicomponent hydrodynamic radiatively driven wind models for A stars
were calculated by Babel (\cite{babela}).
Another radiation driven wind model was calculated by Achmad {\etal}
(\cite{AFGvitr}), but they did not take into account non-LTE effects.
Generally, for extended atmospheres the NLTE effects are stronger.
An attempt to calculate spherically symmetric NLTE model atmospheres with
a stellar wind of Deneb was done by Aufdenberg et al.
(\cite{hauschildt}).
However, as the authors note, they were not able to fit the observed
H$\alpha$ line profile.
Work on developing consistent NLTE wind models of A supergiants is
currently in progress (Krti\v{c}ka \& Kub\'at,
\tentosbornik).


\subsection{Limb darkening}

\begin{figure}
\resizebox{\hsize}{!}
{\includegraphics{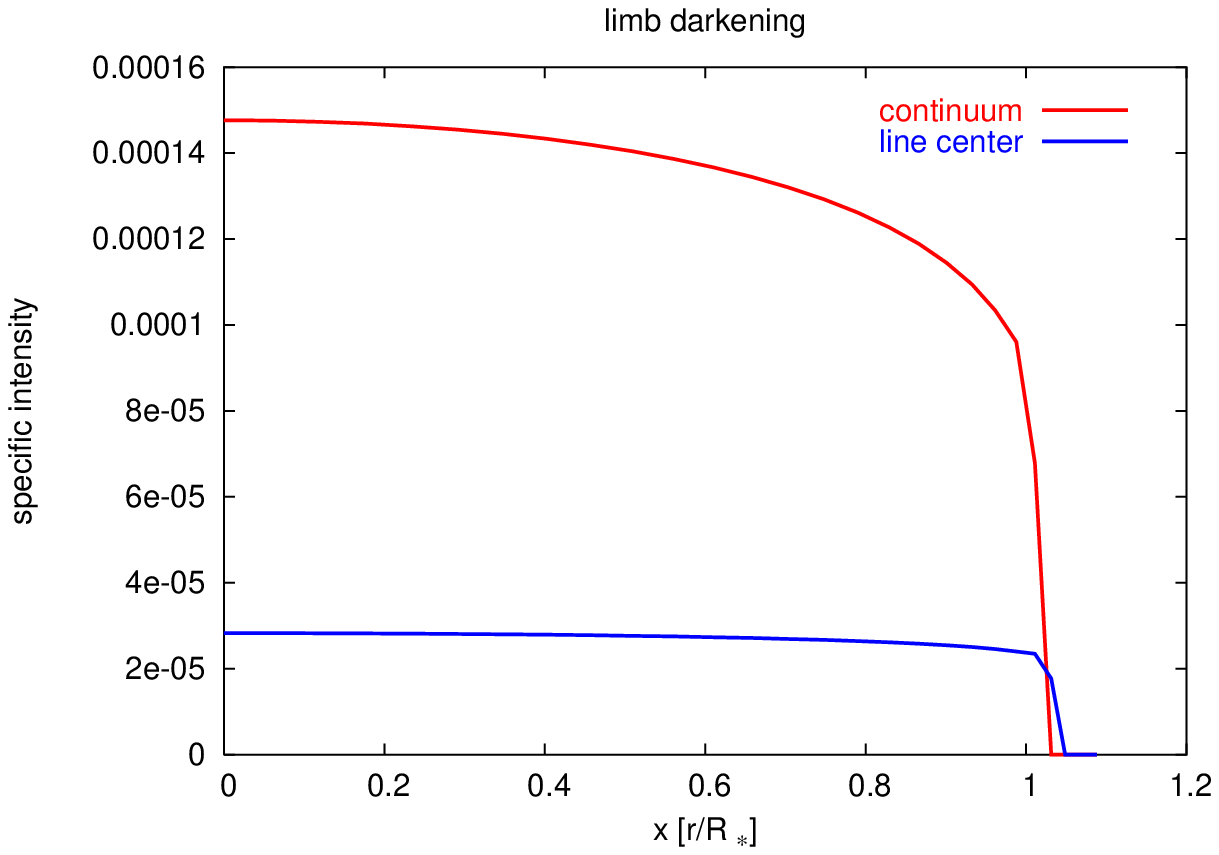}
 \includegraphics{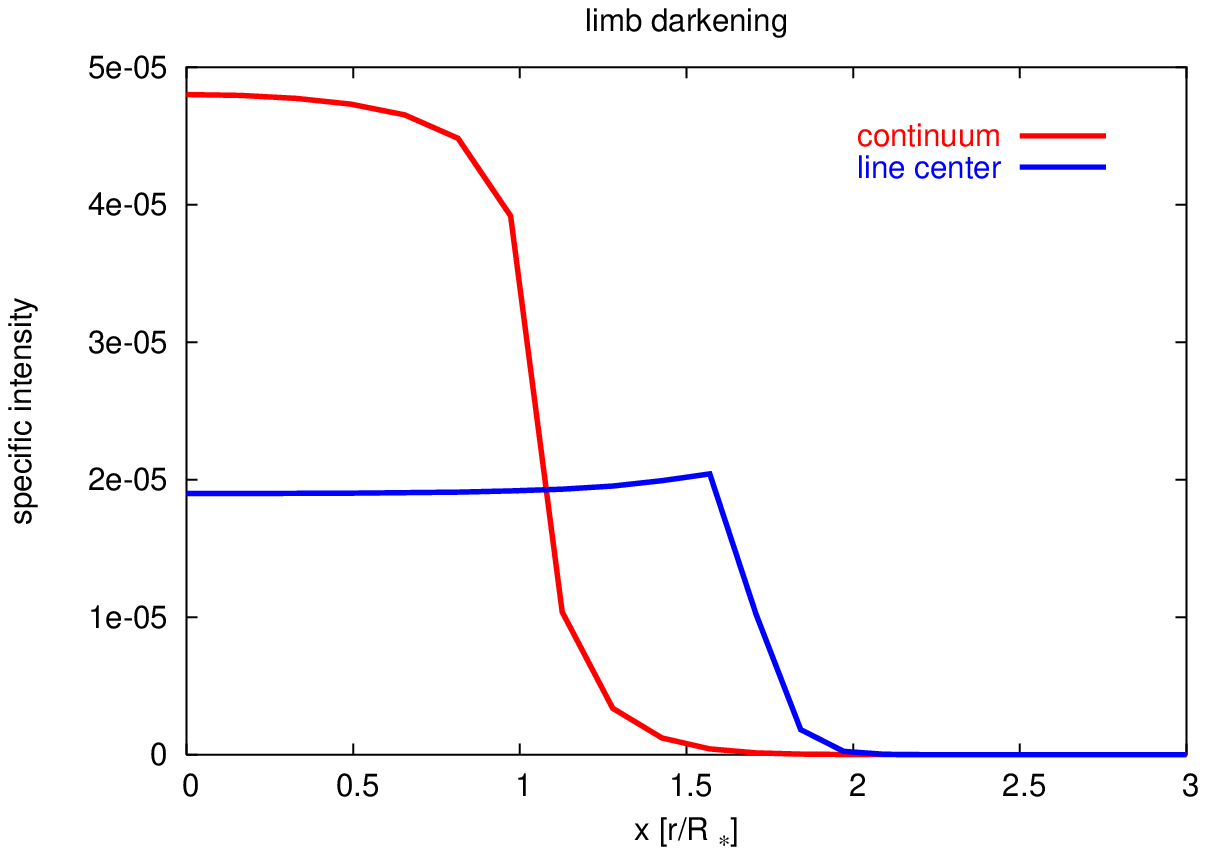}}
\caption{The comparison of limb darkening for the thin atmosphere 
 (left panel) and the extended atmosphere (right panel). These results
 were obtained from the Kor\v{c}\'{a}kov\'{a} \& Kub\'{a}t code (\cite{kk}).}
 \label{okrajtenkytlusty}
\end{figure}

An important property of emergent radiation from stellar atmospheres is
the angle dependence of the specific intensity $I(\theta)$, usually
called limb darkening.
Approximate limb darkening laws are often used
(see, e.g., Allen, \cite{allen} or Gray, \cite{konvoluce}).
These approximate laws don't
describe the angular dependence of the specific intensity
very well even in the case of thin stellar atmospheres.
In addition, limb darkening is strongly frequency dependent (Hadrava \&
Kub\'at \cite{hadku}), and in the center of a line,
even limb brightening instead of darkening may appear, as can be seen in
 Figure\,\ref{okrajtenkytlusty}.
This effect becomes very strong especially for extended
stellar atmospheres.

For a correct description of limb darkening it is necessary to
use model atmospheres with
a better geometry than plane-parallel. 
This has been done by Claret \& Hauschildt (\cite{okraj}), who, using a
spherically symmetric code, calculated limb darkening for a grid of
model atmospheres from A to G spectral types. 

Knowledge of an accurate limb darkening law is necessary for
detecting stellar spots (see Kjurkchieva, \cite{skvrny}).
It is also very important for the analysis of interferometric
measurements, as well as for the correct treatment of stellar
rotation.
Luckily, eclipses from binaries (e.g., Twigg
\cite{zakryt1}), and the gravitational microlensing effect
(Heyrovsk\'y \cite{mikrococky}) measurements of limb darkening 
are possible for at least some of the distant stars.

\subsection{Rotation}

Stellar rotation is usually being accounted by convolving
the rotation profile with the static line profile (see
Gray \cite{konvoluce}).
This technique is inaccurate not only due to using only approximate
limb darkening laws, but also because the dependence of limb darkening
on frequency as well as  gravity darkening are neglected.
A more accurate method is to integrate the static intensity over the
rotating disc of the star, where gravity darkening can also be included.
A ``simple'' way of including gravity darkening in this
integration is described by Collins (\cite{gr.dark}).
This method was used by Gulliver {\etal} (\cite{gulliver})
to determine the rotation of Vega.

Stellar rotation is not like a rigid body rotation, as it is usually
assumed, but it is probably differential.
Until now, there are only few measurements of  differential rotation.
Unfortunately, in many works, the authors use the simplest form of
the limb darkening law and the obtained results can be
used only with reservations.

For an exact description of stellar rotation the plane-parallel
model atmosphere is insufficient.
It is necessary to use a hydrodynamic multidimensional model atmosphere,
at least an axially symmetric one together with the correct description
of the radiation field.
Recently, we started to work on such a model (Kor\v{c}\'akov\'a
{\etal}, \tentosbornik).

\section{Conclusions}

Full NLTE model atmospheres of A stars are difficult to calculate, since
the lines of different ions form at very different depths, 
and therefore there is a need for 
a huge amount of depth points to resolve all line formation
regions sufficiently.
Only LTE model atmospheres together with NLTE calculations for
individual ions have been calculated in significant amounts.
This underlines the need for more intensive calculations of NLTE model
atmospheres of A stars.
Adding consistently new different physical processes to NLTE model
atmospheres calculations, such as diffusion (see Michaud,
{\tentosbornik}), magnetic fields (see Moss, {\tentosbornik}),
and more accurate treatment
of convection zones (see Kupka, {\tentosbornik}), is necessary.
This is the real challenging task the near future.

\begin{acknowledgments}
The authors would like to thank Dr. Ad\'ela Kawka for her comments on
the manuscript.
This research has made an extensive use of the ADS.
This work was supported by grants GA \v{C}R 205/02/0445, 205/04/P224,
and 205/04/1267,
The Astronomical Institute Ond\v{r}ejov is supported
by projects K2043105 and Z1003909.
\end{acknowledgments}

\newcommand{\MRT}[1]{in W. Kalkofen (ed.), \textit{Methods in Radiative
        Transfer}, Cambridge Univ. Press., p.~#1}

\newcommand{\NRT}[1]{in W. Kalkofen (ed.), \textit{Numerical Radiative
        Transfer}, Cambridge Univ. Press, p.~#1}

\newcommand{\SABCM}[1]{in L. Crivellari, I. Hubeny, \& D. G. Hummer
        (eds.), \textit{Stellar Atmospheres: Beyond Classical Models},
        NATO ASI Series C~341, Kluwer Academic Publishers, p.~#1}

\newcommand{\PNPA}[1] {\textit{ASPC} 44, #1}
%

\newcommand{\SAM}[1] {\textit{ASPC} 288, #1}
%

\newcommand{\MSA}[1] {\textit{IAUS} 210, #1}

\end{document}